\documentclass[conference]{IEEEtran}
\IEEEoverridecommandlockouts
\usepackage{relsize}
\usepackage{hhline}
\usepackage{multirow}
\usepackage{algorithm}
\usepackage{algorithmic}
\usepackage{commath}
\usepackage{amsmath,amssymb,amsfonts}
\usepackage{graphicx}
\usepackage{textcomp}
\usepackage{xcolor}
\def\BibTeX{{\rm B\kern-.05em{\sc i\kern-.025em b}\kern-.08em
    T\kern-.1667em\lower.7ex\hbox{E}\kern-.125emX}}
\usepackage{amsthm}
\usepackage{biblatex}
\newtheorem{theorem}{Theorem}
\newtheorem{lemma}[theorem]{Lemma}

\begin{document}

\title{Unsupervised Learning for Identifying High Eigenvector Centrality Nodes: A Graph Neural Network Approach\\
}

\author{
\IEEEauthorblockN{Appan Rakaraddi}
\IEEEauthorblockA{\textit{School of Computer Science,} 
\textit{NTU}\\
Singapore \\
appan001@e.ntu.edu.sg}
\and
\IEEEauthorblockN{Mahardhika Pratama}
\IEEEauthorblockA{\textit{School of Computer Science,} 
\textit{NTU}\\
Singapore \\
mpratama@ntu.edu.sg}
}

\maketitle

\begin{abstract}
 The existing methods to calculate the Eigenvector Centrality(EC) tend to not be robust enough for determination of EC in low time complexity or not well-scalable for large networks, hence rendering them practically unreliable/ computationally expensive. So, it is of the essence to develop a method that is scalable in low computational time. Hence, we propose a deep learning model for the identification of nodes with high Eigenvector Centrality. There have been a few previous works in identifying the high ranked nodes with supervised learning methods, but in real-world cases, the graphs are not labelled and hence deployment of supervised learning methods becomes a hazard and its usage becomes impractical. So, we devise \textbf{CUL}(\textbf{C}entrality with \textbf{U}nsupervised \textbf{L}earning) method to learn the relative EC scores in a network in an unsupervised manner. To achieve this, we develop an Encoder-Decoder based framework that maps the nodes to their respective estimated EC scores. 
%  The Encoder takes the network and its features as inputs and maps the nodes to their respective embedding vectors while preserving the network topology and its other properties. The Decoder transforms the node embedding vectors to their respective estimated EC values.
 Extensive experiments were conducted on different synthetic and real-world networks. We compared CUL against a baseline supervised method for EC estimation similar to some of the past works. It was observed that even with training on a minuscule number of training datasets, CUL delivers a relatively better accuracy score when identifying the higher ranked nodes than its supervised counterpart. We also show that CUL is much faster and has a smaller runtime than the conventional baseline method for EC computation. The code is available at \url{https://github.com/codexhammer/CUL}.
\end{abstract}

\begin{IEEEkeywords}
Eigenvector Centrality, Unsupervised learning, Graph Neural Network
\end{IEEEkeywords}

\section{Introduction}
Graphs are the most prominent and ubiquitous data structures with their presence ranging from road networks, social media networks, pandemic spread evolving network, knowledge graphs to myriad of other places. The mining of the graph based on different metrics hence becomes an extremely important rationale for a thorough analysis of these different networks. Some of the important metrics of a graph are Eigenvector centrality(EC) of the nodes in the graph which delineates a node's influence in the network based on different factors.

For example, consider a social media network like Twitter where the users can be considered as the nodes in the network and the follow requests are the edges between the nodes. A user \textbf{A} with a 1000 followers would seem more influential than a user \textbf{B} with 10 followers, but measuring the influence just on this premise can often be misleading. If each of the 10 followers of user \textbf{B} have followers in the range of millions and if every tweet made by user \textbf{B} is retweeted by his/her followers; and in contrast, if the 1000 followers of user \textbf{A} have very fewer number of followers, then the influence of user \textbf{A} will be significantly less compared to the that of user \textbf{B}. This is where the importance of Eigenvector centrality comes into play for identifying the actual influential nodes rather than just identification of the nodes with the highest number of connections. In other words, nodes with high Eigenvector score are connected to nodes with relatively high scores and nodes which are lower-ranked are connected to nodes with low-valued Eigenvector scores. The Eigenvector Centrality computation by Power Iteration method \cite{newman} is the widely used standardised method. The number of iterations required for convergence is dependent on the network topology and hence the number of iterations $k$ can vary extremely.  So, we need a robust method which can scale up well and still maintain a good accuracy.
So given a graph, we should be able to map each node to its respective EC value. \par

To achieve this, we convert the task of learning the EC values to a node regression problem in an unsupervised learning setting. This can be achieved by an Encoder-Decoder framework that takes in a graph input and maps the nodes to their respective relative EC scores. 
The Encoder generates the low-dimensional embedding matrix of the network by aggregating features from the neighbouring nodes and the Decoder takes the embedding matrix  as input and maps it to estimated EC scores of the nodes. In many real-world scenarios, identification of the top-$\mathcal{N}\%$ of the nodes is crucial compared to determining the EC values of all the nodes in the network. 
In this work, we leverage Graph Neural Networks \cite{gnn} or GNNs to generate the node embedding vector. GNNs have gained a lot traction in the recent years with the application domains ranging from classification tasks \cite{peng2018large,dabhi2020nodenet,errica2020fair}, prediction tasks \cite{li2017diffusion}, Combinatorial Optimisation problems \cite{khalil2017learning,li2018combinatorial,gasse2019exact} etc. 
The embedding matrix can be generated by different node embedding functions like Graph Convolutional Network(GCN) \cite{kipf2016semi}, GraphSAGE \cite{hamilton2018inductive}, Graph Attention Network (GAT) \cite{velivckovic2017graph} etc. We will review some of these methods and also examine how the different aggregation schemes affect over the quality of the embedding and the overall results. This type of Encoder-Decoder framework for Centrality calculation has been formerly introduced to calculate Betweenness Centrality \cite{fan2019learning,maurya2019fast} by extensively training them on smaller graphs and then tested on large real-world networks.
The major shortcoming of these methods is that they are supervised learning methods, making the label generation task as an overhead. This adds to the computational cost that is usually unaccounted for while predicting the centrality values. Hence, we propose a method to predict Eigenvector centrality of the nodes in a network in a completely unsupervised manner and rank the nodes in the order of their centrality values. \par

We introduce \textbf{CUL} (\textbf{C}entrality with \textbf{U}nsupervised   \textbf{L}earning), a GNN based Encoder-Decoder framework for identifies the higher EC nodes in unsupervised manner. The Encoder generates the node representation embedding and the Decoder maps this embedding to its centrality value by training the model in end-to-end manner in unsupervised fashion. \newline

To summarise our contributions:
\begin{itemize}
    \item We propose an Encoder-Decoder based framework and derive an Objective function for unsupervised learning which takes in the graph data and the node features as input and generates the EC scores of the nodes. 
    
    \item We implement different variants of graph neural networks to generate the embedding in the Encoder. We then compare the top-$\mathcal{N}\%$ accuracy variation across these methods. We show that even with a small number of training samples, the model performs with better accuracy than a baseline algorithm.
    
    \item Experiments were conducted on different synthetic and real-world datasets.  We show that the computational times for identification of the top-$\mathcal{N}\%$ are much lower in comparison to the baseline algorithm. The code is available at \url{https://github.com/codexhammer/CUL}. %\url{https://github.com/codexhammer/CUL}.
\end{itemize}

\section{Related Work}

The node regression problem has been one of the prominent applications in the Graph Neural Networks as GNNs can generate high-quality embedding of the network based on the topological structure of the graph. The quality of the embedding can vastly differ based on the variant of the aggregation technique formulated to generate a low-dimensional embedding of size $\mathbb{R}^{n\times d}$ from the higher dimensional adjacency matrix, $A\in \mathbb{R}^{n\times n}$ and the feature vector, $F \in \mathbb{R}^{n\times c}$. \par

There have been different methods proposed for evaluation of centrality using Graph Neural Networks which have demonstrated a far superior result in comparison to the conventional techniques in terms of computational time for very high accuracy in results. \cite{fan2019learning} devised an Encoder-Decoder framework for encoding and mapping a network to the nodes' respective Betweenness Centrality values with a very high accuracy and low computation times compared to other methods for approximating betweenness centrality. \cite{maurya2019fast} proposed approximation for betweenness centrality by calculating the product scores of in-degree and out-degree of the nodes. \cite{grando2018machine} proposed a more generalised method to estimate the 8 different centrality values. It takes two centrality measures to estimate the remaining centrality values of the nodes, specifically eigenvector and degree centrality were chosen assuming their computational cost is low to calculate the high computational cost centrality values. This was accomplished by training on small-scale synthetically generated networks and a neural network was trained with the known centrality measures to predict the unknown centrality values. \cite{Mendonca_2020} improved upon this method by considering only one centrality measure i.e., degree centrality in order to predict the rest of the centrality measures. \par

There has been lot of work done in generating more efficient node embedding based on information aggregating functions
% with 
% (i) Factorisation methods like Laplacian Eigenmaps  \cite{belkin2001laplacian}, HOPE \cite{ou2016asymmetric} etc. (ii) Random-walk based methods like \textit{node2vec} \cite{grover2016node2vec}, DeepWalk \cite{perozzi2014deepwalk} etc. (iii)
Deep Learning based methods  like GCN, GraphSAGE, GAT, DGCNN \cite{wang2019dynamic} etc. In this paper, we will be using GCN, GraphSAGE and GAT to generate the network embedding and will compare the results based on this output. We will be discussing these 3 deep learning methods in the upcoming sections along with how they can be used as an Encoder and also compare the variation in accuracy of the results based on the type of the embedding used.

\section{Preliminaries}

\subsection{Definition}

\textit{Eigenvector centrality}(EC) determines the measure of the influence of a node in a network based on its connections. The nodes in the network which are connected to higher-scoring nodes have a higher EC value or contribute more towards the relative importance of a node in comparison with the equal number of connections to the lower-valued nodes. \par

Let $G=(V,E)$ be a graph with $V$ being the set of $n$ vertices and $E$ being the set of $m$ edges. If $\textbf{A}\in \mathbb{R}^{n \times n}$ is the adjacency matrix with $A_{ij}=1$ if vertex $i$ is connected to vertex $j$, else 0. The feature vector, $\textbf{F} \in \mathbb{R}^{n \times f}$ defines the nodes features which will be used as the model input. This is discussed in the later section.
For an eigenvector \textbf{x}, if $x_i$ is the centrality score of the vertex $i$ and $\lambda $ is the eigenvalue associated with \textbf{x}, then EC is defined by:\par

\begin{align*}
    x_i = \frac{1}{\lambda} \sum_{j=1}^{n}A_{ij}x_j
\end{align*}

This can be written as:

\begin{equation}
    \textbf{A}\textbf{x}=\lambda \textbf{x}
\end{equation}

The EC vector is associated with the largest eigenvalue. By Perron-Frobenius theorem \cite{perron}, since \textbf{A} is a non-negative matrix, there exists a unique largest eigenvalue $\lambda$ which is positive. Hence, a non-negative eigenvector, which is the EC of the matrix also exists. 

\subsection{Graph Embedding}

In this section, we introduce the different embedding techniques that have been used in the Encoder for generating the graph embedding. The Encoder takes the graph and its node features as inputs and encode each of the node into an embedding vector in a latent low-dimensional space using different neighbourhood aggregation schemes for neighbouring node features.

\subsubsection{GCN} The GCN \cite{kipf2016semi} method showed that a simple layer-wise propagation graph neural network can be developed by approximation of first order approximation of the spectral filters on graphs. 

\begin{equation}
    H = \bar{A} F W
\end{equation}

 where $\bar{A} = \hat{D}^{-\frac{1}{2}}\hat{A}\hat{D}^{-\frac{1}{2}}$, $\hat{A} = A+I_n$ is the adjacency matrix with added self loops, $\hat{D}_{ii}=\sum _j \hat{A}_{ij}$ and $F$ is the feature matrix.
\\
\subsubsection{GraphSAGE} The GraphSAGE \cite{hamilton2018inductive} method presented an inductive based scalable approach for node representation based on the graph topology. The methods uses \textit{sample} and \textit{aggregate} where a node's \textit{k}-hop neighbours' features are sampled and aggregated to generate a neighbouring nodes' embedding. This is combined with the node's current embedding to generate a new embedding. 

\begin{align}
\begin{split}
 h^k_{N(v)} &= \text{MEAN}(h_u^{(k-1)},\forall u \in N(v)) 
\\
h^k_v & = \sigma (W^k.([h_v^{(k-1)},h^k_{N(v)}]))
\\
h^k_v & =  \frac{h^k_v}{||h^k_v||_2}
\end{split}
\end{align}

where $h^k_v$ is the hidden representation of node $v$ in the $k$-th iteration and $N(v)$ are the sampled neighbouring nodes of node $v$.
\\
\subsubsection{GAT} The GAT \cite{velivckovic2017graph} method defined the graph attention layer by developing attention \cite{Vaswani} mechanism on graphs and stacking the layers. For a layer, the coefficients of a pair of nodes $(i,j)$ can be expressed as:

\begin{equation}
    \label{eq:attention}
    \alpha _{ij} = \displaystyle \frac{\exp \left(\text{LeakyReLU} \ (\textbf{a}^T [\textbf{W}h_i \ || \  \textbf{W}h_j] ) \right)   }{\sum _{k \in N_i} \exp \left(\text{LeakyReLU} \  (\textbf{a}^T [\textbf{W}h_i \ || \  \textbf{W}h_k] \right) )  }
\end{equation}

where $N_i$ represents the nodes neighbouring to node $i$ and ``$||$'' is the concatenation operation,  $\textbf{h}$ is the feature matrix and $\textbf{W}$ is the trainable weight matrix. The attention mechanism is performed by a single-layer feed forward network parameterized by the weight vector $\textbf{a} \in \mathbb{R}^{2f'}$. The soft-maxed attention coefficients in Equation \ref{eq:attention}, with \textit{K} multi-heads \cite{Vaswani} is used to generate the node feature representation.
% passed through a non-linear activation function $\sigma$ is applied to a linear combination of the features corresponding to the final output features i.e.,

% \begin{equation}
%     h'_i = \sigma \left(\sum _{j \in N_i} \alpha_{ij}\textbf{W} h_j \right)
% \end{equation}

% Multi-head attention \cite{Vaswani} is used to stabilize the learning process with \textit{K}-independent attention mechanisms to compute the hidden states to obtain the output feature representations:

\begin{equation}
        h'_i = \Bigg\Vert_{k=1}^K  \sigma \left(\sum _{j \in N_i} \alpha_{ij}
        \textbf{W} h_j \right)
\end{equation}
\\
\section{CUL}
There are many conventional techniques for obtaining the dominant eigenvector of a network like the Power Iteration, Arnoldi Iteration \cite{Arnoldi1951ThePO}, Jacobi-Davidson method \cite{sleijpen2000jacobi} etc. In our scenario, we will avail the technique behind the Power iteration technique in formulating the loss function. The reasoning behind the use of Power iteration method lies in the feasibility in the formulation of the loss function as a closed form expression which we will derive below in the upcoming sections.

\subsection{Power Iteration}

\begin{figure*}[t]
    \includegraphics[width=\textwidth, height=8cm]{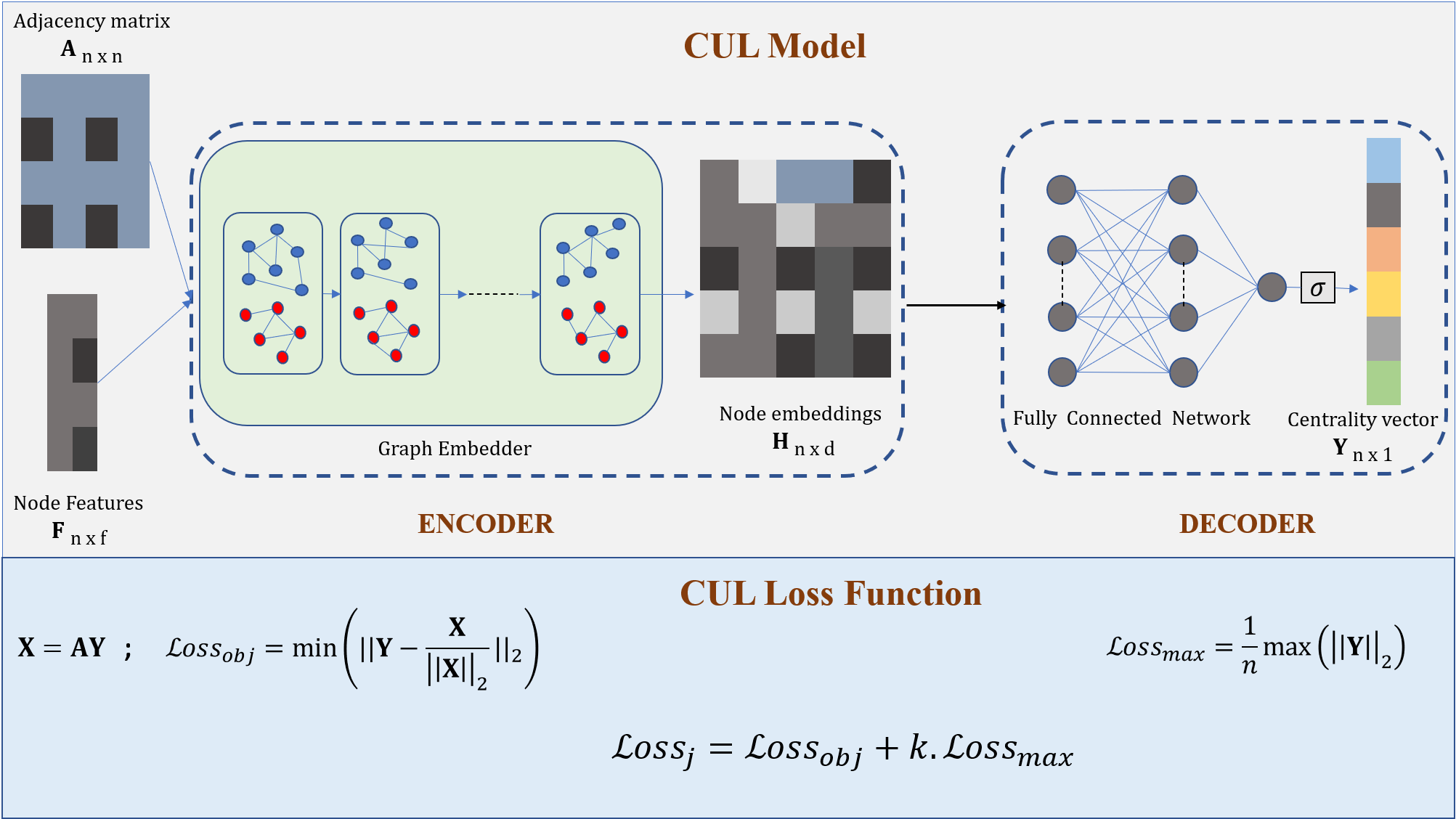}
    \caption{CUL framework for identification of high-valued EC nodes in a network. The joint loss function is the sum of the objective function loss for difference reduction and output maximisation vector.}
    \label{fig:model}
\end{figure*}

The convergence of the eigenvector associated with the largest eigenvalue with the Power iteration method \cite{convergence, panju2011iterative} can be leveraged in modelling the loss function. Formally, the Power iteration is expressed as:

\begin{equation}
\label{eqn:Power}
   \textbf{r}_{k+1} = \frac{\textbf{A}\textbf{r}_k}{||\textbf{A}\textbf{r}_k||_2}
\end{equation}

 where $\textbf{r}_k$ and $\textbf{r}_{k+1}$ are the values of a randomly initialised vector $\textbf{r}_0 \in \mathbb{R}^{n \times 1}$ in the $k$-th and $(k+1)$-th iteration respectively. Then, the vector $\textbf{r}_0$ converges
%  strictly in the presence of a dominant eigenvalue 
  to the eigenvector associated with the highest eigenvalue. This can be easily proved by the standard Power iteration convergence theorem which we have presented in Lemma \ref{lemma:conv}.

\begin{lemma}
\label{lemma:conv}
Let a %diagonalizable
matrix $\textbf{A}\in\mathbb{R}^{n \times n}$ has $n$  eigenvectors i.e., $\{\textbf{v}_1, \textbf{v}_2,..,\textbf{v}_n\}$ corresponding to eigenvalues $\{\lambda _1, \lambda _2,...,\lambda _n\}$. Assuming $\textbf{v}_1$ and $\lambda _1$ are the largest eigenvalue and its corresponding eigenvector respectively, then any random vector $\textbf{r}_0 \in \mathbb{R}^{n \times 1}$ can be expressed as:

\begin{align*}
\textbf{A}^k\textbf{r}_0 \approx c \textbf{v}_1
\end{align*}    

where $k \rightarrow \infty$ for some constant value $c$. \par

\end{lemma}

\begin{proof}

% Let \textbf{V} =  $[\textbf{v}_1^T, \textbf{v}_2^T,..,\textbf{v}_n^T]$ 
% be the matrix with eigenvectors of \textbf{A} as the column vectors and \textbf{D} be the diagonal matrix with principal diagonal populated with the eigenvalues of \textbf{A}. Then by Jordan-canonical form, 

% $$\textbf{A} = \textbf{VDV}^{-1}$$
The random vector $\textbf{r}_0$ can be expressed as a linear combination of \textit{independent} eigenvectors $\{\textbf{v}_1, \textbf{v}_2,..,\textbf{v}_m\}$ as 

$\textbf{r}_0  = c_1\textbf{v}_1 + c_2\textbf{v}_2 + ... +c_m\textbf{v}_m$

$\implies \textbf{Ar}_0 =  c_1\textbf{Av}_1 + c_2\textbf{Av}_2 + ... +c_m\textbf{Av}_m$

$\implies \textbf{Ar}_0 =  c_1\lambda_1\textbf{v}_1 + c_2\lambda_2\textbf{v}_2 + ... +c_n\lambda_m\textbf{v}_m$

Multiplying \textbf{A} repeatedly $k$ times on both sides,

$\implies \textbf{A}^k\textbf{r}_0 =  c_1\lambda_1^k\textbf{v}_1 + c_2\lambda_2^k\textbf{v}_2 + ... +c_m\lambda_m^k\textbf{v}_m$

$\implies \textbf{A}^k\textbf{r}_0 =  \mathlarger{\lambda_1^k\left(c_1\textbf{v}_1 + c_2\left (\frac{\lambda_2}{\lambda_1}\right )^k\textbf{v}_2 + ... +c_m\left(\frac{\lambda_m}{\lambda_1}\right)^k\textbf{v}_m\right )}$

As $k \rightarrow \infty$ and \ $\mathlarger{\norm{\frac{\lambda_i}{\lambda_1}}^k << 1}$ for $i=2,3,...,m$

$\implies \textbf{A}^k\textbf{r}_0 \approx c_1\lambda_1^k\textbf{v}_1 $

$\implies \textbf{A}^k\textbf{r}_0 \approx c\textbf{v}_1 $ where $c$ is a constant.

% From Equation \ref{eqn:Power}, 
% $$\textbf{r}_{k+1} = \frac{\textbf{VDV}^{-1}\textbf{r}_k}{||\textbf{VDV}^{-1}\textbf{r}_k||} = \frac{(\textbf{VDV}^{-1})^{k+1}\textbf{r}_0}{||(\textbf{VDV}^{-1})^{k+1}\textbf{r}_0||}$$

% $$\implies \textbf{r}_{k+1} = \frac{\textbf{V D}^{k+1} \textbf{V}^{-1}\textbf{r}_0}{||\textbf{VD}^{k+1}\textbf{V}^{-1}\textbf{r}_0||}   $$

% $$\implies \textbf{r}_{k+1} = \frac{\textbf{V D}^{k+1} \textbf{V}^{-1} (c_1\textbf{v}_1 + c_2\textbf{v}_2 + ... +c_n\textbf{v}_n)}{||\textbf{VD}^{k+1}\textbf{V}^{-1}(c_1\textbf{v}_1 + c_2\textbf{v}_2 + ... +c_n\textbf{v}_n)||}$$

% $$\implies \textbf{r}_{k+1} = \frac{\textbf{V D}^{k+1}  (c_1\textbf{e}_1 + c_2\textbf{e}_2 + ... +c_n\textbf{e}_n)}{||\textbf{VD}^{k+1}(c_1\textbf{e}_1 + c_2\textbf{e}_2 + ... +c_n\textbf{e}_n)||} \quad (say)$$

\end{proof}

We leverage the above method to derive a suitable loss function for Eigenvector Centrality computation. The loss function is combination of 2 parts: \textit{Objective loss function} and \textit{Maximising loss function}.

\subsection{Objective loss function}

CUL returns an output vector $\textbf{Y} \in \mathbb{R}^{n \times 1}$ vector where $\text{Y}_i$ represents the relative centrality ranking of the node $i$. Based on Equation \ref{eqn:Power}, a vector \textbf{X} is computed in each epoch as shown in Equation \ref{eqn:x}:

\begin{equation}
\label{eqn:x}
    \textbf{X} = \textbf{A}\textbf{Y}
\end{equation}

Hence, the objective loss function can be updated as:

\begin{align*}
    \mathcal{L}oss_{obj} = \sum_{i=1}^n \sqrt{\left(  \text{Y}_i-   \frac{\text{X}_i}{||\textbf{X}||_2}  \right)^2  }
\end{align*}

\begin{equation}
    \mathcal{L}oss_{obj} = \min \left(\left\|\textbf{Y} - \frac{\textbf{X}}{||\textbf{X}||_2}\right\|_2 \right)
\label{eqn:obj}
\end{equation}

where $||.||_2$ denotes the L2 norm of a vector. Note that the minimisation of the loss function is done after the normalization of \textbf{X} to prevent gradient explosion problem that occurs due to the operation performed in Equation \ref{eqn:x} in every epoch.

\subsection{Maximising loss function}

As it can be observed in Equation \ref{eqn:obj},  \textbf{X} in the objective loss function converges towards the trivial solution which is a zero vector, as \textbf{Y} also becomes a zero vector. To avoid this problem and drive the model output vector \textbf{Y} away from the zero vector, we  maximise \textbf{Y} at every epoch to create a moving target to avoid the trivial solution, hence the\textit{ Maximising loss function} is considered. The maximisation of \textbf{Y} is done by considering its absolute value to avoid the zero mean problem.

\begin{align*}
\mathcal{L}oss_{max} = \frac{1}{n}\sqrt{\sum_{i=1}^n \text{Y}_i^2}    
\end{align*}

\begin{equation}
\label{eqn:max}
    \mathcal{L}oss_{max} = \frac{1}{n}\max \left (||\textbf{Y}||_2 \right )
\end{equation}

Hence, Equation \ref{eqn:obj} ensures that the distance between iterative product output \textbf{X} and the model output \textbf{Y} is minimised 
while the vector \textbf{Y} is non-zero which is ensured by Equation \ref{eqn:max}. 

\subsubsection{Joint loss function}

Hence, the effective joint loss function is the summation of the 2 losses:

\begin{align*}
    \mathcal{L}oss_{j} = \mathcal{L}oss_{obj} + k\mathcal{L}oss_{max}
\end{align*}
or
\begin{equation}
    \label{eqn:joint}
    \mathcal{L}oss_{j} = \min \left(\left\|\textbf{Y} - \frac{\textbf{X}}{||\textbf{X}||_2}\right\|_2 \right) + k \left ( \frac{1}{n}\max \left (||\textbf{Y}||_2 \right )   \right )
\end{equation}
 where $k$ is a scaling constant. In this work, we have set the value of $k=1$ as a hyper-parameter.
 
Note that Equation \ref{eqn:max} can also be written as:

\begin{equation*}
    \mathcal{L}oss_{max} = -\frac{1}{n}\min \left (||\textbf{Y}||_2 \right )
\end{equation*}
 
since the maximisation of \textbf{Y} is equivalent to minimisation of -\textbf{Y}. This is essential as the loss function is now minimisable as a whole. Hence Equation \ref{eqn:joint} can be re-written as:

\begin{equation}
    \label{eqn:joint_min}
    \mathcal{L}oss_{j} = \min  \left( \left(\left\|\textbf{Y} - \frac{\textbf{X}}{||\textbf{X}||_2}\right\|_2 \right) - k \left ( \frac{1}{n} \left (||\textbf{Y}||_2 \right )   \right ) \right )
\end{equation}
This loss function is minimized per every minibatch and the iteration in Equation \ref{eqn:x} is computed per epoch.
We also test for different variants of the loss function by considering L1 and L2 norms for loss minimization. This is discussed in the Experimental section.

\subsection{Model Analysis}
The model inputs are node features vector \textbf{F} and the pre-processed network. We describe the details regarding the pre-processing step in the later section. For a node $v$, its initial feature vector is $[d_v]$ where $d_v$ is the degree of the node. 

\subsubsection{Encoder} 

The Encoder aggregates the node features across \textit{K}-hop neighbourhood. In our work, we aggregate till 2-hop neighbours i.e., \textit{K}=2. Hence, the Encoder is a 2-layer graph embedding module connected in  cascade. The training time complexity varies depending on the number of training iterations, number of training samples and the loss convergence. The model during the testing time has a fixed time complexity which is used for the evaluation of the model efficiency.
% But, once the model is trained it can be tested on various graphs and hence this computation time is not considered in the final computation time as it is a one time process. 
The Encoder inputs are defined by $\textbf{Z} = ENC(\textbf{A},\textbf{Q};\ \Theta _{ENC})$ where \textbf{Z} is the Encoder output, \textbf{Q} is the feature vector \textbf{F} in the first layer of the Encoder function with the the output \textbf{Z} being the corresponding next values of \textbf{Q} in the further layers and $\Theta _{ENC}$ is a function which can be GCN, GAT or GraphSAGE that takes \textbf{A} and \textbf{Q} as the inputs.

\subsubsection{Decoder} 

The Decoder is a 4-layer fully connected layer followed by a Leaky ReLU activation function which maps the node embedding to its EC score. This is denoted by $\textbf{Y} = DEC(\textbf{Z}; \ \Theta _{DEC})$ where $\Theta _{DEC}$ is the MLP function which takes \textbf{Z} as inputs. This is described in Algorithm \ref{alg-cul}.

%Add details regarding algo

\subsubsection{Runtime analysis}
In the test phase, the Encoder has a time complexity of $O(|E|)$ for a sparsely connected network as most of the real-world networks fall under this category. The Decoder fully connected network has a time complexity of $O(|V|)$. Thus, the overall time complexity of the model is $O(|V|+|E|)$ for estimation of Eigenvector Centrality.
% It has a time complexity of $O(|V|)$ which is highly parallelisable during the testing phase and this can greatly reduce the estimated total time complexity of $O(|E|+|V|)$ of the model to a much lower estimate. 
The iterative method has a time complexity of $O(k.|E|)$ (\textit{k} is the no. of iterations) with the theoretical maximum time complexity of $O(|V||E|)$ or even worse in case of non-convergence within the error-bound \cite{complexity}. Equation \ref{eqn:joint_min} is minimised at every iteration during the train mode. In our experiments, it was empirically observed that accuracy reached its peak for 150 iterations.

\begin{algorithm}
\caption{CUL algorithm}
\label{alg-cul}
\begin{algorithmic}[1]
\renewcommand{\algorithmicrequire}{\textbf{Input:}}
 \renewcommand{\algorithmicensure}{\textbf{Output:}}
 \REQUIRE Network adjacency matrix \textbf{A}; \vfil \ \quad Features $\{F_v \in \mathbb{R}^f, \ \  \forall v\in V\}$
 \ENSURE  EC value $\{Y_v \in \mathbb{R}, \ \  \forall v\in V\}$ \\

 \begin{center}
     \textit{Encoder}
 \end{center} 
  \STATE Initialise \textbf{Q} = \textbf{F} for the first layer;
  \STATE Set $\Theta _{ENC}$ function to either GCN, GAT or GraphSAGE;
  \FOR{$k=1 \ \text{to} \ K$}
  \STATE $\textbf{Z} \leftarrow \  \Theta _{ENC}(\textbf{A},\textbf{Q})$;
%   \STATE $\textbf{Q} \leftarrow \textbf{Z}$ ; \vfil
%   \STATE $\textbf{Z} \leftarrow \  \Theta _{ENC}(\textbf{A},\textbf{Q})$;
  \STATE $\textbf{Q} \leftarrow \textbf{Z}$
  \ENDFOR
  \begin{center}
      \textit{Decoder}
  \end{center} 
  \STATE $\textbf{Y} \leftarrow \ \Theta _{DEC}(\textbf{Z})$;
  
  \RETURN $\textbf{Y}$
  \STATE Update the joint loss function as defined in Equation \ref{eqn:joint_min} during backpropagation during the training phase in end-to-end manner with Adam optimizer.

\end{algorithmic}
\end{algorithm}

Algorithm \ref{alg-cul} details the flow for each epoch for the joint training of the Encoder-Decoder CUL model.

\section{Results and Analysis}

\subsection{Experiments}
The main comparison standard that we have adapted here to compare our model with is the standard implementation of NetworkX 2.5 algorithm for calculation of Eigenvector Centrality and a supervised training version with the same model architecture.
Particularly, we compare the results based on metrics of top-5\% nodes, top-10\% nodes, top-15\% and top-20 \% of the crucial nodes identification. We describe the datasets used, comparison metrics and other hyper-parameter and hardware configuration settings in the upcoming sections.

\subsubsection{Datasets}

We conducted experiments under different embedding schemes on different real-world and synthetic data sets to prove the efficiency of our model. To demonstrate the applicability of our framework under different graph embedding schemes, we have used a different embedding technique for each of the real-world dataset.  We give details regarding the datasets below:

\subsubsection{Real world datasets}
The below datasets are used as they correspond to the networks where EC tends to be a very important network property.

\begin{itemize}
    \item \textit{cit-DBLP} \cite{dblp} is a citation network dataset. The vertex of the graph depicts a document and the edge depicts if there is a citation between the documents. 
    
    \item \textit{email-Enron} \cite{enron} is an email communication network. The nodes of the network are the email addresses and the edge between node \textit{i} and node \textit{j} is present if at least one email is communicated between \textit{i} and \textit{j}.
    
    \item \textit{com-DBLP} \cite{com} is a co-authorship network. The nodes are the authors and the there is an edge between two authors if they have collaboration on a work together. 
    
    \item \textit{rt-retweet-crawl} \cite{retweet} is crawled from the \textit{Twitter} network. The nodes are the Twitter users and the edges denote the retweets between the users.
    
\end{itemize}
The properties of the network are specified in Table \ref{tab:dataset}.

\begin{figure*}[t]
    \centering
    \includegraphics[scale=0.3]{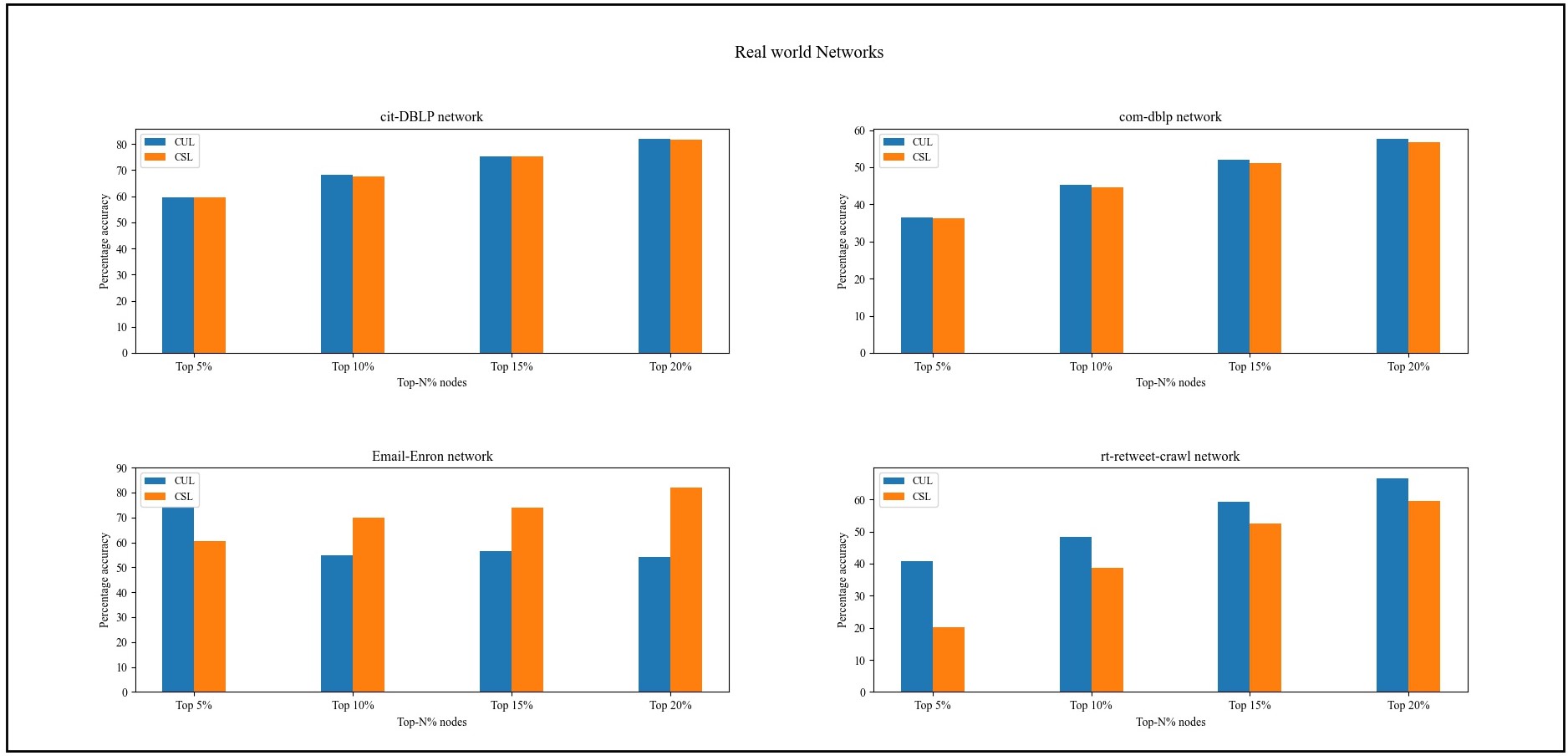}
    \caption{Top-$\mathcal{N}$\% accuracy comparison between CUL and CSL on the real world networks. CUL outperforms CSL in most of the cases.}
    \label{fig:real}
\end{figure*}

\subsubsection{Synthetic datasets}
We have used three different synthetically generated graphs for training and testing our model, namely:

\begin{itemize}
    \item \textit{Scale-free(SF) graphs}: We used NetworkX 2.5 to generate 50 Scale-Free graphs with n=1000 for training purposes. For testing, Scale-Free graphs with size: 10,000, 20,000, 50,000 and 100,000 nodes were generated.
    
    \item \textit{Barabasi-Albert(BA) graphs}: We used NetworkX 2.5 to generate 50 Barabasi-Albert graphs with n=1000 and m=4 for training purposes. For testing, Barabasi-Albert graphs with size: 10,000, 20,000, 50,000 and 100,000 nodes and m=4 were generated.
    
    \item \textit{Powerlaw cluster(PL) graphs}: We used NetworkX 2.5 to generate 50 Barabasi-Albert graphs with n=1000, m=4 and p=0.05 for training purposes. For testing, Powerlaw cluster graphs with size: 10,000, 20,000, 50,000 and 100,000 nodes with m=4 and p=0.05 were generated.
    
\end{itemize}

\begin{table}[H]
\centering
\caption{Real world dataset specifics}
\label{tab:dataset}
\resizebox{\linewidth}{!}{%
\begin{tabular}{lllcc} 
\hline
\textbf{Network} & \multicolumn{1}{c}{\textbf{\textbar{}V\textbar{}}} & \multicolumn{1}{c}{\textbf{\textbar{}E\textbar{}}} & \textbf{Average clustering coefficient} & \multicolumn{1}{l}{\textbf{Diameter}} \\ 
\hline
cit-DBLP & 12,591 & 49,635 & 0.01 & 10 \\
Email-Enron & 36,692 & 183,831 & 0.49 & 11 \\
com-dblp & 317,080 & 1,049,866 & 0.63 & 21 \\
rt-retweet-crawl & 1,112,702 & 2,278,852 & 0.01 & N.A. \\
\hline
\end{tabular}
}
\end{table}

\subsection{TSNE analysis}

\begin{figure*}[t]
    \centering
    \includegraphics[width=\textwidth, height=3.5cm]{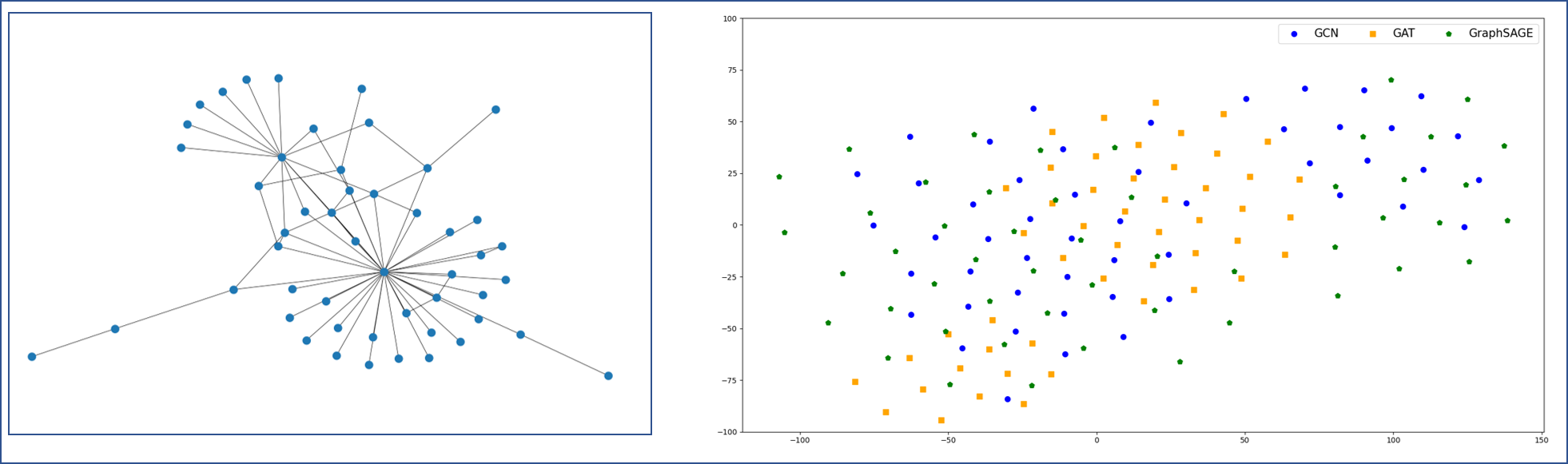}
    \caption{TSNE analysis of the network is done with GCN, GAT and GraphSAGE embedding schemes. GCN and GraphSAGE projection points are more spread apart giving a better accuracy in comparison to closely-knit points of GAT.}
    \label{fig:tsne}
\end{figure*}

It has been previously shown that with even an untrained model and for randomly initialized weights, GCN produced an embedding that closely resembled the community structure of a network. For our analysis, we used the TSNE plot to observe the embedding of a network on a 2D plane. We generated a 50 node Scale-Free network and trained the model in end-to-end manner. The Encoder embedding were projected on the 2D plane to observe how well the nodes with similar EC values are clustered together for the embedding schemes: GCN, GAT and GraphSAGE. Based on the TSNE plot in Figure \ref{fig:tsne}, GCN and GraphSAGE are more widely spread apart  making it easier to obtain a better accuracy in identifying the high-valued EC nodes, compared to GAT which produces more closely knit-together structure producing lower a relatively lower accuracy. This has also been reflected in Table \ref{tab:sf}, Table \ref{tab:ba} and Table \ref{tab:pl} where GCN, GraphSAGE have relatively better accuracy in comparison to GAT on different synthetic networks.

\subsection{Baseline methods and comparison metrics}

\subsubsection{Choice of baseline methods}

There are different methods for estimation of Eigenvector Centrality like Power iteration \cite{convergence}, Lanczos algorithm \cite{lanczos-1950}, Inverse iteration \cite{ipsen-1997}, Rayleigh quotient iteration \cite{parlett-1974} etc. But, the major drawbacks of these methods lies in their computational difficulties. For example, the QR algorithm for computation has a cubic convergence rate for eigenvector estimation making the iterations computationally expensive. Inverse iteration method relies on on solving linear equations in the matrix form leading to high space complexity which rapidly increases the storage space for larger matrices. A similar problem is also experienced with Lanczos algorithm. So, based on the theoretical complexities for comparison, we opt for the suitable method as the baseline that can be practically scaled for large scale matrices. Power iteration method for eigenvector computation has proven to be the most suitable method for large-scale sparse matrices where the explicit storage of coefficient matrix consumes a lot of hardware memory space \cite{google-iter}. As such, since our method focuses on the scalability factor, we deploy Power iteration method as one of our baselines.

\subsubsection{Baseline methods}

We  now outline the baseline methods against which CUL will be tested for performance comparison.

\begin{enumerate}
    \item  \textbf{C}entrality with \textbf{S}upervised \textbf{L}earning or \textbf{CSL}: This has the same model architecture as CUL, but it is a supervised learning method. We first label the nodes of the synthetically generated graphs with their respective Eigenvector Centrality values and then train it in a  supervised learning way. We compare the accuracy  of CUL against this method to evaluate the performance as the testing runtimes will be similar. Note that this method is quite similar to the \cite{fan2019learning} (it is used to estimate Betweenness Centrality) where they define an Encoder similar to GraphSAGE and the Decoder is a MLP network which is also a supervised learning method albeit with some differences like the number of layers, loss function etc. We didn't find any deep learning model specifically for Eigenvector centrality estimation, hence we configured the CUL model to train in supervised environment as a baseline comparison.
    
    \item \textbf{Iterative}: We use the iterative method for calculation of Eigenvector Centrality. This is implemented via NetworkX 2.5 algorithm for calculation of Eigenvector centrality of a graph. We compare the runtime of CUL against this method.
\end{enumerate}

As mentioned earlier, we  demonstrate that the CUL performs better in terms of accuracy during the testing phase in comparison to its supervised counterpart i.e., CSL when the same number of training data samples are used for both of them. 

\par 
For performance comparison, we identify the top-$\mathcal{N}\%$ of the nodes using different graph embedding techniques for a given network where top-$\mathcal{N}\%$ is computed for  top-5\%, top-10\%, top-15\% and top-20\%.

\small $$\text{Top-}\mathcal{N}\% = \frac{|\{ \text{Returned top -} \mathcal{N}\% \text{ nodes}\} \cap \{ \text{True top -}\mathcal{N}\% \text{ nodes} \}|}{ \lceil |\text{V}| \times \mathcal{N}\%  \rceil}$$

\normalsize
where $|$V$|$ is the number of nodes in a network $\lceil x \rceil$ is the ceiling function \cite{fan2019learning}.

\subsection{Other Settings}

The model is implemented in PyTorch with Adam optimizer with learning rate of 0.001 running on Tesla T4 GPU on the Google Colab cloud. For the Encoder, 2 embedding layers of the same type are used back to back i.e.,
$ENC(\textbf{A},ENC(\textbf{A},\textbf{F};\Theta _{ENC});\Theta _{ENC})$. In our experiments, we use Batch Gradient Descent with the embedding dimension of each layer set to 128. The implementation of all the embedding techniques have been done using geometric deep learning library: Pytorch Geometric \cite{pytorch}. Note that the actual implementation is done with the network stored in the edge list format as adjacency matrix takes a lot of redundant memory space, thus making it impractical to for larger datasets.  The network pre-processing that was earlier mentioned in the Model Analysis section is done so as to ensure that the input network is converted into appropriate edge index format for loading into Pytorch Geometric module. The edge index model also reduces the total memory space for storing the dataset compared to the adjacency matrix.

\subsection{Evaluation on Synthetic Datasets}

\begin{table*}
\centering
\caption{Scale Free graph accuracy comparison with mean and standard deviation}
\label{tab:sf}
\resizebox{\linewidth}{!}{%
\begin{tabular}{|l|ll|ll|ll|ll|ll|ll|ll|ll|ll|ll|ll|ll|} 
\hline
\multicolumn{1}{|c|}{\multirow{3}{*}{\begin{tabular}[c]{@{}c@{}}SF \\ graph size\end{tabular}}} & \multicolumn{8}{c|}{\textbf{GCN}} & \multicolumn{8}{c|}{\textbf{GAT}} & \multicolumn{8}{c|}{\textbf{GraphSAGE}} \\ 
\cline{2-25}
\multicolumn{1}{|c|}{} & \multicolumn{2}{c|}{Top-5\%} & \multicolumn{2}{c|}{Top-10\%} & \multicolumn{2}{c|}{Top-15\%} & \multicolumn{2}{c|}{Top-20\%} & \multicolumn{2}{c|}{Top-5\%} & \multicolumn{2}{c|}{Top-10\%} & \multicolumn{2}{c|}{Top-15\%} & \multicolumn{2}{c|}{Top-20\%} & \multicolumn{2}{c|}{Top-5\%} & \multicolumn{2}{c|}{Top-10\%} & \multicolumn{2}{c|}{Top-15\%} & \multicolumn{2}{c|}{Top-20\%} \\ 
\cline{2-25}
\multicolumn{1}{|c|}{} & \multicolumn{1}{c}{\textbf{CUL }} & \multicolumn{1}{c|}{\textbf{CSL }} & \multicolumn{1}{c}{\textbf{CUL }} & \multicolumn{1}{c|}{\textbf{CSL }} & \multicolumn{1}{c}{\textbf{CUL }} & \multicolumn{1}{c|}{\textbf{CSL }} & \multicolumn{1}{c}{\textbf{CUL }} & \multicolumn{1}{c|}{\textbf{CSL }} & \multicolumn{1}{c}{\textbf{CUL }} & \multicolumn{1}{c|}{\textbf{CSL }} & \multicolumn{1}{c}{\textbf{CUL }} & \multicolumn{1}{c|}{\textbf{CSL }} & \multicolumn{1}{c}{\textbf{CUL }} & \multicolumn{1}{c|}{\textbf{CSL }} & \multicolumn{1}{c}{\textbf{CUL }} & \multicolumn{1}{c|}{\textbf{CSL }} & \multicolumn{1}{c}{\textbf{CUL }} & \multicolumn{1}{c|}{\textbf{CSL }} & \multicolumn{1}{c}{\textbf{CUL }} & \multicolumn{1}{c|}{\textbf{CSL }} & \multicolumn{1}{c}{\textbf{CUL }} & \multicolumn{1}{c|}{\textbf{CSL }} & \multicolumn{1}{c}{\textbf{CUL }} & \multicolumn{1}{c|}{\textbf{CSL }} \\ 
\hline
\textbf{1000 } & 73.0$\pm$5.3 & \textbf{81.3$\pm$1.4 } & 77.8$\pm$4.9 & \textbf{78.1$\pm$6.3 } & 83.7$\pm$4.3 & \textbf{84.6$\pm$4.1 } & 70.2$\pm$2.9 & \textbf{74.8$\pm$4.1 } & \textbf{67.3$\pm$2.9 } & 62.8$\pm$0.9 & \textbf{69.0$\pm$5.6 } & 56.2$\pm$4.5 & \textbf{71.7$\pm$4.7 } & 54.0$\pm$4.4 & \textbf{72.2$\pm$3.5 } & 57.5$\pm$2.3 & 71.4$\pm$5.9 & \textbf{82.6$\pm$1.4 } & 78.2$\pm$3.9 & \textbf{78.8$\pm$6.1 } & 82.6$\pm$5.6 & \textbf{84.6$\pm$4.0 } & 77.3$\pm$7.5 & \textbf{76.6$\pm$4.0 } \\
\textbf{10000 } & 72.2$\pm$6.2 & \textbf{75.6$\pm$2.0 } & 73.1$\pm$13.0 & \textbf{78.8$\pm$0.8 } & 69.2$\pm$6.1 & \textbf{76.5$\pm$6.8 } & 61.5$\pm$5.2 & \textbf{69.1$\pm$10.4 } & \textbf{59.6$\pm$0.8 } & 50.4$\pm$0.7 & \textbf{62.2$\pm$1.2 } & 50.0$\pm$1.0 & \textbf{68.3$\pm$0.2 } & 52.8$\pm$0.3 & \textbf{74.7$\pm$2.7 } & 53.4$\pm$1.7 & 70.5$\pm$1.5 & \textbf{78.6$\pm$4.0 } & 81.5$\pm$1.7 & \textbf{81.9$\pm$3.4 } & \textbf{81.7$\pm$2.1 } & 77.3$\pm$7.0 & \textbf{76.7$\pm$5.1 } & 67.2$\pm$7.8 \\
\textbf{20000 } & 65.8$\pm$2.9 & \textbf{68.3$\pm$2.6 } & 62.5$\pm$15.6 & \textbf{63.2$\pm$14.6 } & \textbf{57.3$\pm$1.4 } & 56.5$\pm$9.0 & 56.2$\pm$0.1 & \textbf{65.6$\pm$6.1 } & \textbf{71.2$\pm$9.2 } & 48.8$\pm$1.5 & \textbf{75.8$\pm$6.5 } & 50.2$\pm$1.3 & \textbf{76.2$\pm$2.6 } & 51.9$\pm$1.4 & \textbf{81.4$\pm$3.7 } & 49.7$\pm$0.6 & 66.6$\pm$0.0 & \textbf{69.2$\pm$2.6 } & \textbf{71.2$\pm$0.9 } & 63.7$\pm$15.3 & \textbf{70.9$\pm$0.7 } & 57.4$\pm$10.1 & \textbf{68.5$\pm$2.1 } & 66.4$\pm$5.7 \\
\textbf{50000 } & \textbf{67.3$\pm$0.0} & 67.1$\pm$0.8 & 61.3$\pm$16.5 & \textbf{61.5$\pm$14.7 } & 55.0$\pm$2.3 & \textbf{61.7$\pm$6.9 } & 65.1$\pm$20.9 & \textbf{71.1$\pm$19.9 } & \textbf{58.1$\pm$1.0 } & 45.1$\pm$0.4 & \textbf{68.8$\pm$0.2 } & 47.7$\pm$1.4 & \textbf{77.1$\pm$1.0 } & 47.2$\pm$2.2 & \textbf{84.0$\pm$1.8 } & 45.3$\pm$3.2 & \textbf{71.4$\pm$3.4 } & 67.4$\pm$1.2 & \textbf{79.2$\pm$0.8 } & 61.3$\pm$14.9 & \textbf{68.9$\pm$0.9 } & 59.6$\pm$7.7 & \textbf{73.8$\pm$0.8 } & 66.9$\pm$19.1 \\
\textbf{100000 } & 49.9$\pm$3.7 & \textbf{60.2$\pm$1.6 } & 47.4$\pm$3.0 & \textbf{48.2$\pm$2.3 } & 41.9$\pm$5.7 & \textbf{43.5$\pm$4.1 } & \textbf{68.5$\pm$2.5 } & 64.9$\pm$11.7 & \textbf{54.6$\pm$1.3 } & 43.6$\pm$1.4 & \textbf{67.3$\pm$5.0 } & 47.6$\pm$0.8 & \textbf{75.0$\pm$6.7 } & 47.8$\pm$1.5 & \textbf{79.8$\pm$8.4 } & 46.1$\pm$2.0 & 58.3$\pm$0.0 & \textbf{60.9$\pm$1.0 } & \textbf{57.0$\pm$0.0 } & 46.0$\pm$2.3 & \textbf{64.3$\pm$0.0 } & 42.3$\pm$4.7 & \textbf{84.3$\pm$0.0 } & 64.9$\pm$12.7 \\
\hline
\end{tabular}
}
\end{table*}

\begin{table*}
\centering
\caption{Barabasi-Albert graph accuracy comparison with mean and standard deviation}
\label{tab:ba}
\resizebox{\linewidth}{!}{%
\begin{tabular}{|l|ll|ll|ll|ll|ll|ll|ll|ll|ll|ll|ll|ll|} 
\hline
\multicolumn{1}{|c|}{\multirow{3}{*}{\begin{tabular}[c]{@{}c@{}}BA\\ graph size\end{tabular}}} & \multicolumn{8}{c|}{\textbf{GCN}} & \multicolumn{8}{c|}{\textbf{GAT}} & \multicolumn{8}{c|}{\textbf{GraphSAGE}} \\ 
\cline{2-25}
\multicolumn{1}{|c|}{} & \multicolumn{2}{c|}{Top-5\%} & \multicolumn{2}{c|}{Top-10\%} & \multicolumn{2}{c|}{Top-15\%} & \multicolumn{2}{c|}{Top-20\%} & \multicolumn{2}{c|}{Top-5\%} & \multicolumn{2}{c|}{Top-10\%} & \multicolumn{2}{c|}{Top-15\%} & \multicolumn{2}{c|}{Top-20\%} & \multicolumn{2}{c|}{Top-5\%} & \multicolumn{2}{c|}{Top-10\%} & \multicolumn{2}{c|}{Top-15\%} & \multicolumn{2}{c|}{Top-20\%} \\ 
\cline{2-25}
\multicolumn{1}{|c|}{} & \multicolumn{1}{c}{\textbf{CUL }} & \multicolumn{1}{c|}{\textbf{CSL }} & \multicolumn{1}{c}{\textbf{CUL }} & \multicolumn{1}{c|}{\textbf{CSL }} & \multicolumn{1}{c}{\textbf{CUL }} & \multicolumn{1}{c|}{\textbf{CSL }} & \multicolumn{1}{c}{\textbf{CUL }} & \multicolumn{1}{c|}{\textbf{CSL }} & \multicolumn{1}{c}{\textbf{CUL }} & \multicolumn{1}{c|}{\textbf{CSL }} & \multicolumn{1}{c}{\textbf{CUL }} & \multicolumn{1}{c|}{\textbf{CSL }} & \multicolumn{1}{c}{\textbf{CUL }} & \multicolumn{1}{c|}{\textbf{CSL }} & \multicolumn{1}{c}{\textbf{CUL }} & \multicolumn{1}{c|}{\textbf{CSL }} & \multicolumn{1}{c}{\textbf{CUL }} & \multicolumn{1}{c|}{\textbf{CSL }} & \multicolumn{1}{c}{\textbf{CUL }} & \multicolumn{1}{c|}{\textbf{CSL }} & \multicolumn{1}{c}{\textbf{CUL }} & \multicolumn{1}{c|}{\textbf{CSL }} & \multicolumn{1}{c}{\textbf{CUL }} & \multicolumn{1}{c|}{\textbf{CSL }} \\ 
\hline
\textbf{1000 } & 62.5$\pm$2.5 & \textbf{72.4$\pm$3.8 } & 63.0$\pm$4.3 & \textbf{70.2$\pm$3.6 } & 68.1$\pm$3.7 & \textbf{70.2$\pm$2.6 } & 70.7$\pm$2.4 & \textbf{72.1$\pm$2.7 } & 28.0$\pm$0.0 & \textbf{82.6$\pm$1.3 } & 37.0$\pm$0.0 & \textbf{73.2$\pm$3.3 } & 48.6$\pm$0.0 & \textbf{71.0$\pm$2.3 } & 53.0$\pm$0.0 & \textbf{70.4$\pm$2.6 } & 69.5$\pm$2.5 & \textbf{73.6$\pm$3.2 } & \textbf{72.7$\pm$3.6 } & 70.2$\pm$3.6 & \textbf{71.6$\pm$3.7 } & 70.8$\pm$1.8 & \textbf{72.0$\pm$1.9 } & 71.5$\pm$3.7 \\
\textbf{10000 } & \textbf{79.2$\pm$2.0 } & 63.6$\pm$4.0 & \textbf{67.6$\pm$4.9 } & 61.8$\pm$0.9 & \textbf{69.3$\pm$5.2 } & 64.6$\pm$7.0 & \textbf{73.8$\pm$2.1 } & 67.0$\pm$9.6 & \textbf{66.4$\pm$2.8 } & 54.4$\pm$1.6 & \textbf{64.7$\pm$1.3 } & 49.4$\pm$2.0 & \textbf{68.7$\pm$1.4 } & 47.4$\pm$2.5 & \textbf{71.3$\pm$1.6 } & 50.8$\pm$2.2 & \textbf{80.9$\pm$0.4 } & 57.0$\pm$7.0 & \textbf{71.0$\pm$4.8 } & 68.9$\pm$3.6 & \textbf{70.3$\pm$6.3 } & 68.1$\pm$2.7 & \textbf{70.6$\pm$3.3 } & 69.5$\pm$2.5 \\
\textbf{20000 } & \textbf{67.6$\pm$4.0 } & 65.3$\pm$0.9 & \textbf{67.6$\pm$2.9 } & 61.0$\pm$0.2 & \textbf{73.0$\pm$0.3 } & 61.8$\pm$1.1 & \textbf{73.7$\pm$0.5 } & 61.8$\pm$3.4 & \textbf{65.7$\pm$0.9 } & 43.1$\pm$1.3 & \textbf{74.0$\pm$0.7 } & 43.1$\pm$1.3 & \textbf{72.3$\pm$2.3 } & 45.7$\pm$1.5 & \textbf{74.2$\pm$0.2 } & 49.0$\pm$0.8 & \textbf{73.1$\pm$0.8 } & 63.7$\pm$2.3 & \textbf{71.4$\pm$2.9 } & 61.1$\pm$0.2 & \textbf{71.6$\pm$2.5 } & 60.9$\pm$1.5 & \textbf{71.5$\pm$1.7 } & 60.5$\pm$3.4 \\
\textbf{50000 } & \textbf{71.2$\pm$4.0 } & 62.7$\pm$4.1 & \textbf{71.3$\pm$4.3 } & 53.2$\pm$4.2 & \textbf{72.5$\pm$2.2 } & 54.8$\pm$1.3 & \textbf{75.0$\pm$1.3 } & 55.1$\pm$0.3 & \textbf{71.5$\pm$2.4 } & 35.4$\pm$1.1 & \textbf{74.8$\pm$4.9 } & 39.2$\pm$1.4 & \textbf{76.7$\pm$3.8 } & 45.3$\pm$0.5 & \textbf{78.5$\pm$2.2 } & 48.8$\pm$0.4 & \textbf{74.0$\pm$6.5 } & 62.1$\pm$4.7 & \textbf{74.8$\pm$5.1 } & 60.5$\pm$2.9 & \textbf{71.6$\pm$2.5 } & 57.4$\pm$1.1 & \textbf{70.7$\pm$0.5 } & 57.9$\pm$2.3 \\
\textbf{100000 } & \textbf{72.9$\pm$5.7 } & 53.9$\pm$5.0 & \textbf{68.6$\pm$1.8 } & 51.4$\pm$7.8 & \textbf{72.3$\pm$2.7 } & 51.3$\pm$4.6 & \textbf{74.6$\pm$1.7 } & 54.3$\pm$4.2 & \textbf{74.2$\pm$5.2 } & 32.5$\pm$0.9 & \textbf{76.2$\pm$2.1 } & 37.0$\pm$0.8 & \textbf{79.7$\pm$1.3 } & 43.5$\pm$0.7 & \textbf{80.3$\pm$1.5 } & 48.0$\pm$0.9 & \textbf{74.5$\pm$6.5 } & 53.9$\pm$5.3 & \textbf{69.5$\pm$0.4 } & 51.4$\pm$7.8 & \textbf{68.4$\pm$0.3 } & 51.3$\pm$4.6 & \textbf{69.2$\pm$1.3 } & 54.3$\pm$4.2 \\
\hline
\end{tabular}
}
\end{table*}

\begin{table*}
\centering
\caption{Powerlaw cluster graph accuracy comparison with mean and standard deviation}
\label{tab:pl}
\resizebox{\linewidth}{!}{%
\begin{tabular}{|l|ll|ll|ll|ll|ll|ll|ll|ll|ll|ll|ll|ll|} 
\hline
\multicolumn{1}{|c|}{\multirow{3}{*}{\begin{tabular}[c]{@{}c@{}}PL\\ graph size\end{tabular}}} & \multicolumn{8}{c|}{\textbf{GCN}} & \multicolumn{8}{c|}{\textbf{GAT}} & \multicolumn{8}{c|}{\textbf{GraphSAGE}} \\ 
\cline{2-25}
\multicolumn{1}{|c|}{} & \multicolumn{2}{c|}{Top-5\%} & \multicolumn{2}{c|}{Top-10\%} & \multicolumn{2}{c|}{Top-15\%} & \multicolumn{2}{c|}{Top-20\%} & \multicolumn{2}{c|}{Top-5\%} & \multicolumn{2}{c|}{Top-10\%} & \multicolumn{2}{c|}{Top-15\%} & \multicolumn{2}{c|}{Top-20\%} & \multicolumn{2}{c|}{Top-5\%} & \multicolumn{2}{c|}{Top-10\%} & \multicolumn{2}{c|}{Top-15\%} & \multicolumn{2}{c|}{Top-20\%} \\ 
\cline{2-25}
\multicolumn{1}{|c|}{} & \multicolumn{1}{c}{\textbf{CUL }} & \multicolumn{1}{c|}{\textbf{CSL }} & \multicolumn{1}{c}{\textbf{CUL }} & \multicolumn{1}{c|}{\textbf{CSL }} & \multicolumn{1}{c}{\textbf{CUL }} & \multicolumn{1}{c|}{\textbf{CSL }} & \multicolumn{1}{c}{\textbf{CUL }} & \multicolumn{1}{c|}{\textbf{CSL }} & \multicolumn{1}{c}{\textbf{CUL }} & \multicolumn{1}{c|}{\textbf{CSL }} & \multicolumn{1}{c}{\textbf{CUL }} & \multicolumn{1}{c|}{\textbf{CSL }} & \multicolumn{1}{c}{\textbf{CUL }} & \multicolumn{1}{c|}{\textbf{CSL }} & \multicolumn{1}{c}{\textbf{CUL }} & \multicolumn{1}{c|}{\textbf{CSL }} & \multicolumn{1}{c}{\textbf{CUL }} & \multicolumn{1}{c|}{\textbf{CSL }} & \multicolumn{1}{c}{\textbf{CUL }} & \multicolumn{1}{c|}{\textbf{CSL }} & \multicolumn{1}{c}{\textbf{CUL }} & \multicolumn{1}{c|}{\textbf{CSL }} & \multicolumn{1}{c}{\textbf{CUL }} & \multicolumn{1}{c|}{\textbf{CSL }} \\ 
\hline
\textbf{1000 } & 59.0$\pm$9.0 & \textbf{67.7$\pm$2.2 } & 54.5$\pm$1.5 & \textbf{66.5$\pm$2.9 } & 61.0$\pm$1.0 & \textbf{66.9$\pm$2.5 } & 69.7$\pm$1.7 & \textbf{69.8$\pm$2.6 } & 42.0$\pm$0.0 & \textbf{83.1$\pm$3.6 } & 48.0$\pm$0.0 & \textbf{75.0$\pm$4.2 } & 54.0$\pm$0.0 & \textbf{70.4$\pm$3.2 } & 59.0$\pm$0.0 & \textbf{70.1$\pm$1.9 } & 65.3$\pm$1.8 & \textbf{68.4$\pm$2.3 } & 62.6$\pm$1.2 & \textbf{66.6$\pm$2.0 } & 65.5$\pm$0.3 & \textbf{68.4$\pm$1.1 } & 68.8$\pm$3.0 & \textbf{70.4$\pm$2.9 } \\
\textbf{10000 } & \textbf{75.4$\pm$0.0 } & 69.7$\pm$13.3 & 74.2$\pm$0.0 & \textbf{81.6$\pm$1.7 } & 67.7$\pm$0.0 & \textbf{70.9$\pm$0.8 } & 64.6$\pm$0.0 & \textbf{71.3$\pm$2.9 } & \textbf{70.6$\pm$0.0 } & 53.8$\pm$2.8 & \textbf{66.4$\pm$0.0 } & 46.4$\pm$3.3 & \textbf{64.2$\pm$0.0 } & 47.3$\pm$1.9 & \textbf{69.8$\pm$0.0 } & 49.2$\pm$0.8 & 68.7$\pm$12.7 & \textbf{69.7$\pm$13.3 } & 79.9$\pm$2.3 & \textbf{81.7$\pm$1.7 } & \textbf{70.7$\pm$1.3 } & 70.9$\pm$0.8 & \textbf{71.7$\pm$4.5 } & 71.3$\pm$2.9 \\
\textbf{20000 } & 67.9$\pm$0.1 & \textbf{77.9$\pm$2.1 } & 69.7$\pm$1.0 & \textbf{71.7$\pm$5.3 } & 70.7$\pm$0.1 & \textbf{71.4$\pm$3.4 } & \textbf{70.9$\pm$1.5 } & 70.6$\pm$3.2 & \textbf{60.8$\pm$0.5 } & 42.1$\pm$0.4 & \textbf{70.7$\pm$3.1 } & 41.0$\pm$1.8 & \textbf{71.7$\pm$0.9 } & 45.9$\pm$0.7 & \textbf{75.2$\pm$0.1 } & 49.6$\pm$1.4 & 75.7$\pm$1.8 & \textbf{77.0$\pm$1.5 } & \textbf{75.7$\pm$5.3 } & 74.6$\pm$2.3 & \textbf{75.0$\pm$0.6 } & 73.0$\pm$2.2 & \textbf{72.6$\pm$1.8 } & 72.4$\pm$1.0 \\
\textbf{50000 } & 74.7$\pm$6.9 & \textbf{80.0$\pm$6.8 } & 73.5$\pm$2.7 & \textbf{76.8$\pm$1.7 } & 72.8$\pm$2.6 & \textbf{73.9$\pm$2.1 } & \textbf{74.2$\pm$0.6 } & 71.6$\pm$1.1 & \textbf{69.3$\pm$3.8 } & 35.6$\pm$0.8 & \textbf{74.1$\pm$3.5 } & 37.7$\pm$0.6 & \textbf{76.1$\pm$1.6 } & 42.6$\pm$0.6 & \textbf{77.9$\pm$1.1 } & 47.8$\pm$0.1 & 79.1$\pm$7.8 & \textbf{80.0$\pm$6.0 } & 76.8$\pm$1.9 & \textbf{76.8$\pm$1.0 } & 73.2$\pm$1.9 & \textbf{73.9$\pm$2.0 } & 70.9$\pm$0.5 & \textbf{71.6$\pm$1.1 } \\
\textbf{100000 } & \textbf{60.0$\pm$0.0 } & 56.4$\pm$5.7 & \textbf{67.4$\pm$0.0 } & 55.9$\pm$4.3 & \textbf{66.3$\pm$0.0 } & 61.1$\pm$2.2 & \textbf{65.1$\pm$0.0 } & 63.6$\pm$2.1 & \textbf{68.1$\pm$3.4 } & 31.0$\pm$0.1 & \textbf{72.2$\pm$0.6 } & 37.7$\pm$0.6 & \textbf{77.1$\pm$0.2 } & 42.2$\pm$1.1 & \textbf{79.4$\pm$0.2 } & 46.5$\pm$0.9 & 58.3$\pm$2.4 & \textbf{58.7$\pm$3.4 } & \textbf{65.4$\pm$0.9 } & 60.9$\pm$0.6 & \textbf{69.2$\pm$2.6 } & 61.5$\pm$1.9 & \textbf{72.6$\pm$0.6 } & 62.0$\pm$3.7 \\
\hline
\end{tabular}
}
\end{table*}

\begin{table*}
\centering
\caption{Time comparison (in seconds) of Scale Free graph between CUL and Iterative method}
\label{tab:sf-time}
\resizebox{\linewidth}{!}{%
\begin{tabular}{|l|ll|ll|ll|} 
\hline
\multicolumn{1}{|c|}{\multirow{2}{*}{\begin{tabular}[c]{@{}c@{}}SF graph \\ size\end{tabular}}} & \multicolumn{2}{c|}{\textbf{GCN}} & \multicolumn{2}{c|}{\textbf{GAT}} & \multicolumn{2}{c|}{\textbf{GraphSAGE}} \\ 
\cline{2-7}
\multicolumn{1}{|c|}{} & \multicolumn{1}{c}{\textbf{CUL }} & \multicolumn{1}{c|}{\textbf{Iterative }} & \multicolumn{1}{c}{\textbf{CUL }} & \multicolumn{1}{c|}{\textbf{Iterative }} & \multicolumn{1}{c}{\textbf{CUL }} & \multicolumn{1}{c|}{\textbf{Iterative }} \\ 
\hline
\textbf{1000 } & \textbf{0.002$\pm$1.9$\times 10^{-4}$} & 0.011$\pm$1.8$\times 10^{-3}$ & \textbf{0.003$\pm$0.3$\times 10^{-3}$} & 0.012$\pm$2.0$\times 10^{-3}$ & \textbf{0.002$\pm$2.0$\times 10^{-4}$} & 0.018$\pm$3.0$\times 10^{-3}$ \\
\textbf{10000 } & \textbf{0.002$\pm$0.2$\times 10^{-4}$} & 0.177$\pm$0.6$\times 10^{-3}$ & \textbf{0.003$\pm$0.0$\times 10^{-2}$} & 0.091$\pm$8.0$\times 10^{-3}$ & \textbf{0.002$\pm$0.0$\times 10^{-2}$} & 0.091$\pm$6.0$\times 10^{-3}$ \\
\textbf{20000 } & \textbf{0.003$\pm$0.6$\times 10^{-4}$} & 0.323$\pm$9.0$\times 10^{-3}$ & \textbf{0.004$\pm$2.0$\times 10^{-4}$} & 0.217$\pm$7.0$\times 10^{-2}$ & \textbf{0.002$\pm$0.0$\times 10^{-2}$} & 0.244$\pm$5.0$\times 10^{-3}$ \\
\textbf{50000 } & \textbf{0.005$\pm$0.2$\times 10^{-4}$} & 0.713$\pm$2.6$\times 10^{-2}$ & \textbf{0.007$\pm$0.0$\times 10^{-2}$} & 0.776$\pm$1.9$\times 10^{-2}$ & \textbf{0.003$\pm$3.0$\times 10^{-4}$} & 0.713$\pm$2.6$\times 10^{-2}$ \\
\textbf{100000 } & \textbf{0.009$\pm$0.0$\times 10^{-2}$} & 1.622$\pm$5.0$\times 10^{-2}$ & \textbf{0.011$\pm$0.0$\times 10^{-2}$} & 1.412$\pm$1.2$\times 10^{-1}$ & \textbf{0.005$\pm$0.0$\times 10^{-2}$} & 1.478$\pm$0.0$\times 10^{-2}$ \\
\hline
\end{tabular}
}
\end{table*}

\begin{table*}
\centering
\caption{Time comparison (in seconds) of Barabasi-Albert graph between CUL and Iterative method}
\label{tab:ba-time}
\resizebox{\linewidth}{!}{%
\begin{tabular}{|l|ll|ll|ll|} 
\hline
\multicolumn{1}{|c|}{\multirow{2}{*}{\begin{tabular}[c]{@{}c@{}}BA graph \\ size\end{tabular}}} & \multicolumn{2}{c|}{\textbf{GCN}} & \multicolumn{2}{c|}{\textbf{GAT}} & \multicolumn{2}{c|}{\textbf{GraphSAGE}} \\ 
\cline{2-7}
\multicolumn{1}{|c|}{} & \multicolumn{1}{c}{\textbf{CUL }} & \multicolumn{1}{c|}{\textbf{Iterative }} & \multicolumn{1}{c}{\textbf{CUL }} & \multicolumn{1}{c|}{\textbf{Iterative }} & \multicolumn{1}{c}{\textbf{CUL }} & \multicolumn{1}{c|}{\textbf{Iterative }} \\ 
\hline
\textbf{1000 } & \textbf{0.002$\pm$4.0$\times 10^{-4}$} & 0.019$\pm$2.9$\times 10^{-3}$ & \textbf{0.002$\pm$0.0$\times 10^{-4}$} & 0.017$\pm$0.0$\times 10^{-4}$ & \textbf{0.002$\pm$2.3$\times 10^{-4}$} & 0.021$\pm$2.2$\times 10^{-3}$ \\
\textbf{10000 } & \textbf{0.003$\pm$0.0$\times 10^{-4}$} & 0.318$\pm$1.4$\times 10^{-3}$ & \textbf{0.003$\pm$1.1$\times 10^{-4}$} & 0.229$\pm$7.7$\times 10^{-2}$ & \textbf{0.002$\pm$0.5$\times 10^{-4}$} & 0.145$\pm$4.1$\times 10^{-3}$ \\
\textbf{20000 } & \textbf{0.004$\pm$0.0$\times 10^{-4}$} & 0.505$\pm$6.4$\times 10^{-3}$ & \textbf{0.004$\pm$0.0$\times 10^{-4}$} & 0.480$\pm$0.4$\times 10^{-4}$ & \textbf{0.002$\pm$0.3$\times 10^{-4}$} & 0.495$\pm$3.5$\times 10^{-3}$ \\
\textbf{50000 } & \textbf{0.007$\pm$7.0$\times 10^{-4}$} & 1.347$\pm$1.9$\times 10^{-2}$ & \textbf{0.009$\pm$2.4$\times 10^{-4}$} & 1.334$\pm$2.9$\times 10^{-2}$ & \textbf{0.003$\pm$1.4$\times 10^{-4}$} & 1.349$\pm$5.8$\times 10^{-3}$ \\
\textbf{100000 } & \textbf{0.012$\pm$0.0$\times 10^{-4}$} & 2.975$\pm$1.4$\times 10^{-2}$ & \textbf{0.017$\pm$3.8$\times 10^{-4}$} & 2.926$\pm$8.3$\times 10^{-3}$ & \textbf{0.003$\pm$0.3$\times 10^{-4}$} & 2.742$\pm$15.8$\times 10^{-2}$ \\
\hline
\end{tabular}
}
\end{table*}

\begin{table*}
\centering
\caption{Time comparison (in seconds) of Powerlaw graph between CUL and Iterative method}
\label{tab:pl-time}
\resizebox{\linewidth}{!}{%
\begin{tabular}{|l|ll|ll|ll|} 
\hline
\multicolumn{1}{|c|}{\multirow{2}{*}{\begin{tabular}[c]{@{}c@{}}PL graph \\ size\end{tabular}}} & \multicolumn{2}{c|}{\textbf{GCN}} & \multicolumn{2}{c|}{\textbf{GAT}} & \multicolumn{2}{c|}{\textbf{GraphSAGE}} \\ 
\cline{2-7}
\multicolumn{1}{|c|}{} & \multicolumn{1}{c}{\textbf{CUL }} & \multicolumn{1}{c|}{\textbf{Iterative }} & \multicolumn{1}{c}{\textbf{CUL }} & \multicolumn{1}{c|}{\textbf{Iterative }} & \multicolumn{1}{c}{\textbf{CUL }} & \multicolumn{1}{c|}{\textbf{Iterative }} \\ 
\hline
\textbf{1000 } & \textbf{0.003$\pm$4.3$\times 10^{-4}$} & 0.017$\pm$9.4$\times 10^{-4}$ & \textbf{0.002$\pm$0.0$\times 10^{-4}$} & 0.016$\pm$0.0$\times 10^{-4}$ & \textbf{0.002$\pm$0.8$\times 10^{-4}$} & 0.018$\pm$2.6$\times 10^{-3}$ \\
\textbf{10000 } & \textbf{0.003$\pm$0.0$\times 10^{-4}$} & 0.150$\pm$0.0$\times 10^{-4}$ & \textbf{0.003$\pm$0.0$\times 10^{-4}$} & 0.155$\pm$0.0$\times 10^{-4}$ & \textbf{0.002$\pm$0.3$\times 10^{-4}$} & 0.319$\pm$6.6$\times 10^{-3}$ \\
\textbf{20000 } & \textbf{0.004$\pm$0.4$\times 10^{-4}$} & 0.328$\pm$8.5$\times 10^{-3}$ & \textbf{0.332$\pm$1.5$\times 10^{-4}$} & 4.002$\pm$8.5$\times 10^{-2}$ & \textbf{0.002$\pm$0.2$\times 10^{-4}$} & 0.329$\pm$3.4$\times 10^{-3}$ \\
\textbf{50000 } & \textbf{0.007$\pm$0.5$\times 10^{-4}$} & 1.377$\pm$2.2$\times 10^{-2}$ & \textbf{0.009$\pm$0.7$\times 10^{-4}$} & 1.260$\pm$1.2$\times 10^{-2}$ & \textbf{0.003$\pm$0.9$\times 10^{-4}$} & 1.425$\pm$6.3$\times 10^{-3}$ \\
\textbf{100000 } & \textbf{0.012$\pm$0.0$\times 10^{-4}$} & 3.059$\pm$0.0$\times 10^{-4}$ & \textbf{0.017$\pm$2.6$\times 10^{-4}$} & 2.884$\pm$6.7$\times 10^{-2}$ & \textbf{0.003$\pm$6.3$\times 10^{-4}$} & 2.941$\pm$5.0$\times 10^{-2}$ \\
\hline
\end{tabular}
}
\end{table*}

\begin{figure*}[t]
    \centering
    \includegraphics[width=\textwidth, height=6cm]{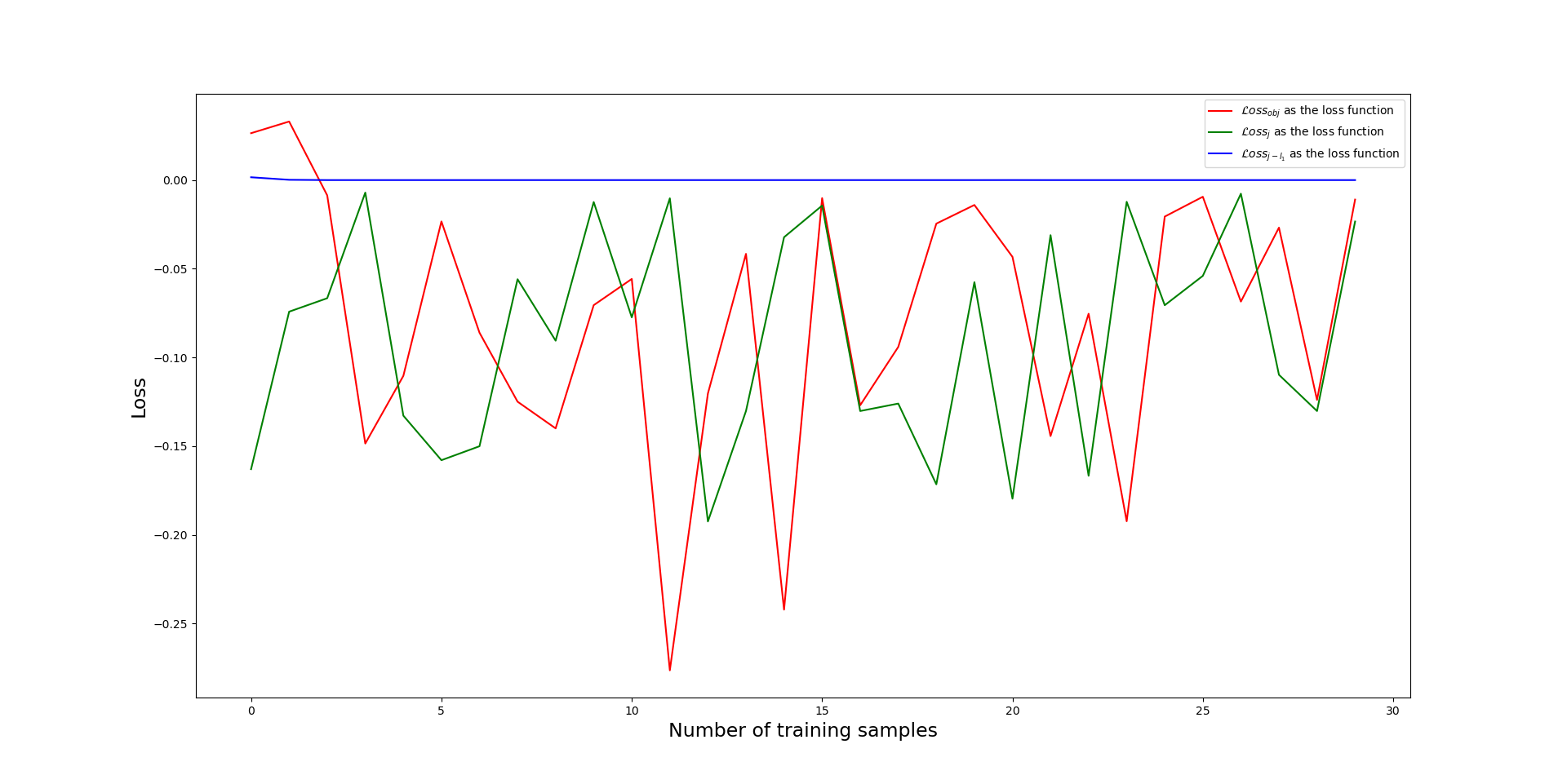}
    \caption{Loss variation vs training samples across different loss functions.}
    \label{fig:loss-epoch}
\end{figure*}

For evaluation purposes, 10 random networks of sizes 10000, 20000, 50000 and 5 random networks of size 100000 are generated of SF, BA and PL type networks. These networks are compared with CSL for accuracy comparison. \par
As shown in Table \ref{tab:sf} for Scale Free networks, CSL didn't scale up well with appropriately in comparison to CUL. But, CSL tend to fare better compared to CUL when ran on smaller scale graphs with nodes less than 50000. This trend is also quite varying depending on the embedding scheme that has been employed. For example, when using GAT, CUL persistently fared better in terms of top-$\mathcal{N}\%$ accuracy whereas with GCN embedding, CUL fared the worst relative to CSL. 
This similar trend was also observed for the BA(Table \ref{tab:ba}) and PL(Table \ref{tab:pl}) graphs, although the top-$\mathcal{N}\%$ accuracy results weren't as skewed in favour of one particular type of embedding. The time comparison is not specified between the two methods as in the testing phase, both the methods are ran on the same model. \par

For time comparison, we bet it against the iterative convergence method for calculation of Eigenvector centrality. Although the iterative method converges to a more accurate value, but CUL is atleast 50-100 times faster in  terms of computation speed (as shown in Table \ref{tab:sf-time}, Table \ref{tab:ba-time}, Table \ref{tab:pl-time}) in each of the different embedding techniques which very well makes up for the slight drop in accuracy. Note that since the features of a node in the graph is its degree, it is assumed that the node degrees are pre-computed during the testing phase. It can be seen that CUL generalises and scales up on par with its supervised counterpart CSL with different embedding techniques even with a small training set of 50 graphs of 1000 nodes.

\subsection{ Evaluation on Real-world Datasets}

The trained model is tested on real-world networks by training on smaller graphs. CUL and CSL were compared across multiple embedding schemes.
\begin{table}[H]
\centering
\caption{Accuracy comparison for real-world dataset}
\label{tab:real}
\resizebox{\linewidth}{!}{%
\begin{tabular}{|l|l|ll|ll|ll|ll|} 
\hline
\multicolumn{1}{|c|}{\multirow{2}{*}{\textbf{Real-world network}}} & \multicolumn{1}{c|}{\multirow{2}{*}{\textbf{Embedding type}}} & \multicolumn{2}{c|}{\textbf{Top-5\%}} & \multicolumn{2}{c|}{\textbf{Top-10\%}} & \multicolumn{2}{c|}{\textbf{Top-15\%}} & \multicolumn{2}{c|}{\textbf{Top-20\%}} \\ 
\cline{3-10}
\multicolumn{1}{|c|}{} & \multicolumn{1}{c|}{} & CUL & CSL & CUL & CSL & CUL & CSL & CUL & CSL \\ 
\hline
cit-DBLP & GraphSAGE & \textbf{59.45 } & 59.45 & \textbf{68.14 } & 67.67 & \textbf{75.21 } & 75.21 & \textbf{81.89 } & 81.57 \\
com-dblp & GAT & \textbf{36.50 } & 36.25 & \textbf{45.20 } & 44.54 & \textbf{52.07 } & 51.20 & \textbf{57.56 } & 56.82 \\
Email-Enron & GCN & \textbf{74.42 } & 60.41 & 54.78 & \textbf{69.89 } & 56.49 & \textbf{73.88 } & 54.04 & \textbf{82.00 } \\
rt-retweet-crawl & GCN & \textbf{40.73 } & 20.06 & \textbf{48.27 } & 38.69 & \textbf{59.14 } & 52.40 & \textbf{66.58 } & 59.47 \\
\hline
\end{tabular}
}
\end{table}
As shown in Table \ref{tab:real}, CUL outperformed CSL when the training environment was set similar. For a slight-drop in accuracy, CUL reduces the computational time with rapid scale-up which provides a strong motivation for deploying CUL. We also performed a statistical test called Mann-Whitney U test \cite{mann-whit} to verify that on average CUL has a higher probability of a more accurate value compared to CSL. Mann-Whiteley U test determines the probability of value selection from 2 independent and non-normal groups with different mean rank values.
% Note that Mann-Whiteley U test is used to determine the probability of a value selected from a group with larger mean rank is greater than a value selected from the other group where the 2 groups are independent, non-normally distributed, categorical classes (here the 2 groups being CUL and CSL).

\begin{table}[H]
\centering
\caption{Time comparison (in seconds) of real-world dataset}
\label{tab:real-time}
\resizebox{\linewidth}{!}{%
\begin{tabular}{|l|l|l|l|} 
\hline
\textbf{Real-world graph} & \textbf{Embedding type} & \multicolumn{1}{c|}{\textbf{Net time}} & \multicolumn{1}{c|}{\textbf{Real time}} \\ 
\hline
cit-DBLP & GraphSAGE & \textbf{0.003} & 0.191 \\
com-dblp & GAT & \textbf{0.069} & 7.127 \\
Email-Enron & GCN & \textbf{0.006} & 1.022 \\
rt-retweet-crawl & GCN & \textbf{0.662} & 26.405 \\
\hline
\end{tabular}
}
\end{table}

\subsection{Ablation study of loss function}

Since the model has a joint loss function, this section analyses the impact of how the loss varies, specifically due to the presence and absence of $\mathcal{L}oss_{max}$ in the joint loss function $\mathcal{L}oss_j$. Hence, we conduct  a study where the second term in the joint loss function is eliminated and the joint loss solely being $\mathcal{L}oss_{obj}$(Refer Table \ref{tab:l2}).

\begin{table}[H]
\centering
\caption{Accuracy on real-world dataset with $\mathcal{L}oss_{obj}$ as the loss function}
\label{tab:l2}
\resizebox{\linewidth}{!}{%
\begin{tabular}{|l|l|l|l|l|l|} 
\hline
\textbf{Real-world network} & \textbf{Embedding type} & \textbf{Top-5\%} & \textbf{Top-10\%} & \textbf{Top-15\%} & \textbf{Top-20\%} \\ 
\hline
cit-DBLP & GraphSAGE & 11.92 & 22.16 & 31.03 & 32.60 \\
com-dblp & GAT & 34.98 & 42.19 & 48.69 & 53.85 \\
Email-Enron & GCN & 7.79 & 10.19 & 11.04 & 12.12 \\
rt-retweet-crawl & GCN & 9.83 & 14.45 & 18.29 & 24.44 \\
\hline
\end{tabular}
}
\end{table}

We also conduct a study by replacing $\mathcal{L}oss_{obj}$ with L1-norm loss function to compare the loss convergence differences between $1^{\text{st}}$-degree norm and $2^{\text{nd}}$-degree norm (i.e., L1-norm and and L2-norm losses). The accuracy values for this tested on the real-world dataset are shown in Table \ref{tab:l1}. Let's say this loss function as $\mathcal{L}oss_{j-l_1}$.

\begin{table}[H]
\centering
\caption{Accuracy on real-world dataset with $\mathcal{L}oss_{j-l_1}$ as the loss function} 
\label{tab:l1}
\resizebox{\linewidth}{!}{%
\begin{tabular}{|l|l|l|l|l|l|} 
\hline
\textbf{Real-world network} & \textbf{Embedding type} & \textbf{Top-5\%} & \textbf{Top-10\%} & \textbf{Top-15\%} & \textbf{Top-20\%} \\ 
\hline
cit-DBLP & GraphSAGE & 33.39 & 38.04 & 42.21 & 46.38 \\
com-dblp & GAT & 36.08 & 45.13 & 50.25 & 55.39 \\
Email-Enron & GCN & 35.94 & 55.92 & 61.59 & 61.9 \\
rt-retweet-crawl & GCN & 39.08 & 29.30 & 25.72 & 24.46 \\
\hline
\end{tabular}
}
\end{table}

During the training process, it was observed that for $\mathcal{L}oss_{obj}$ as the loss function, the output vector \textbf{Y} of the model was driven towards the zero vector, hence with almost no swing in the loss function leading to a constant zero-valued loss. For $\mathcal{L}oss_{j}$ and $\mathcal{L}oss_{j-l_1}$ as the loss function, the presence of the $\mathcal{L}oss_{max}$ term prevented the zero-output vector with a decaying crest-trough type of swing in the loss function. This is shown in Figure \ref{fig:loss-epoch}.

\section{Conclusion and Future Work}

In this work, we discussed about the application of different graph neural networks in an Encoder-Decoder format to predict the Eigenvector Centrality in a completely unsupervised manner on different synthetic and real-world networks. We also showed that CUL provided a good accuracy even when trained with a small training set of 50 graphs. An extension of the work lies in calculation of variants of Eigenvector Centrality like PageRank and Katz Centrality in unsupervised manner. PageRank considers the probability transition matrix (which can be exploited for forming the objective function) for calculation of the centrality of a node. So, an extension of the work can be generalising model architecture for EC variants (like PageRank, Katz Centrality etc.) with only variation in the Objective function.

\printbibliography
\end{document}